\documentstyle[prl,aps,floats,psfig,tabularx]{revtex}
\draft
\newcommand{\KS}{K_S}

\newcommand{\KSTB}{\overline{K}^{*0}}
\newcommand{\LMB}{\Lambda^0}
\newcommand{\DP}{D^+}
\newcommand{\DS}{D^+_s}
\newcommand{\ppp}{\KS \pi^+\pi^+\pi^-}
\newcommand{\ws}{\KS K^-\pi^+\pi^+}
\newcommand{\rs}{\KS K^+\pi^+\pi^-}
\newcommand{\kkp}{\KS K^+K^-\pi^+}

\newcommand{\Ckov}{\v{C}erenkov~}

\newcommand{\RAW}{\rightarrow}

\newcommand{\ETAL}{{\em et al.},~}

\def\nim#1#2#3  {{\it Nucl. Instr. Meth.} {\bf#1}, #2 (#3). }
\def\np#1#2#3   {{\it Nucl. Phys.} {\bf#1}, #2 (#3). }
\def\pcps#1#2#3 {{\it Proc. Cam. Phil. Soc.} {\bf#1}, #2 (#3). }
\def\pl#1#2#3   {{\it Phys. Lett.} {\bf#1}, #2 (#3). }
\def\plc#1#2#3   {{\it Phys. Lett.} {\bf#1}, #2 (#3); }
\def\prep#1#2#3 {{\it Phys. Rep.} {\bf#1}, #2 (#3). }
\def\prev#1#2#3 {{\it Phys. Rev.} {\bf#1}, #2 (#3). }
\def\prevv#1#2#3 {{\it Phys. Rev.} {\bf#1}, #2 (#3). }
\def\prl#1#2#3  {{\it Phys. Rev. Lett.} {\bf#1}, #2 (#3). }
\def\prs#1#2#3  {{\it Proc. Roy. Soc.} {\bf#1}, #2 (#3). }
\def\rmp#1#2#3  {{\it Rev. Mod. Phys.} {\bf#1}, #2 (#3). }
\def\rpp#1#2#3  {{\it Rep. Prog. Phys.} {\bf#1}, #2 (#3). }
\def\zp#1#2#3   {{\it Z. Phys.} {\bf#1}, #2 (#3). }
\def\zpp#1#2#3   {{\it Z. Phys.} {\bf#1}, #2 (#3). }
\def\epj#1#2#3   {{\it Eur. Phys. Jour.} {\bf#1}, #2 (#3). }

\begin{document}

\wideabs{
\title{A measurement of the branching ratios of $\DP$ and $\DS$
hadronic decays to four-body final states containing a $\KS$} 

\author{J.~M.~Link,$^{1}$ M.~Reyes,$^{1,}$\cite{a} P.~M.~Yager,$^{1}$
J.~C.~Anjos,$^{2}$ I.~Bediaga,$^{2}$ C.~G\"obel,$^{2,}$\cite{b}
J.~Magnin,$^{2}$ A.~Massafferi,$^{2}$ J.~M.~de~Miranda,$^{2}$
I.~M.~Pepe,$^{2,}$\cite{c} A.~C.~dos~Reis,$^{2}$
F.~R.~A.~Sim\~ao,$^{2}$ S.~Carrillo,$^{3}$ E.~Casimiro,$^{3,}$\cite{d}
A~S\'anchez-Hern\'andez,$^{3}$ C.~Uribe,$^{3,}$\cite{e}
F.~V\'azquez,$^{3}$ L.~Cinquini,$^{4,}$\cite{f} J.~P.~Cumalat,$^{4}$
B.~O'Reilly,$^{4}$ J.~E.~Ramirez,$^{4}$
E.~W.~Vaandering,$^{4,}$\cite{g} J.~N.~Butler,$^{5}$
H.~W.~K.~Cheung,$^{5}$ I.~Gaines,$^{5}$ P.~H.~Garbincius,$^{5}$
L.~A.~Garren,$^{5}$ E.~Gottschalk,$^{5}$ P.~H.~Kasper,$^{5}$
A.~E.~Kreymer,$^{5}$ R.~Kutschke,$^{5}$ S.~Bianco,$^{6}$
F.~L.~Fabbri,$^{6}$ S.~Sarwar,$^{6}$ A.~Zallo,$^{6}$
C.~Cawlfield,$^{7}$ D.~Y.~Kim,$^{7}$ A.~Rahimi,$^{7}$ J.~Wiss,$^{7}$
R.~Gardner,$^{8}$ Y.~S.~Chung,$^{9}$ J.~S.~Kang,$^{9}$ B.~R.~Ko,$^{9}$
J.~W.~Kwak,$^{9}$ K.~B.~Lee,$^{9,}$\cite{h} H.~Park,$^{9,}$\cite{i}
G.~Alimonti,$^{10}$ M.~Boschini,$^{10}$ B.~Caccianiga,$^{10}$
P.~D'Angelo,$^{10}$ M.~DiCorato,$^{10}$ P.~Dini,$^{10}$
M.~Giammarchi,$^{10}$ P.~Inzani,$^{10}$ F.~Leveraro,$^{10}$
S.~Malvezzi,$^{10}$ D.~Menasce,$^{10}$ M.~Mezzadri,$^{10}$
L.~Milazzo,$^{10}$ L.~Moroni,$^{10}$ D.~Pedrini,$^{10}$
C.~Pontoglio,$^{10}$ F.~Prelz,$^{10}$ M.~Rovere,$^{10}$
A.~Sala,$^{10}$ S.~Sala,$^{10}$ T.~F.~Davenport~III,$^{11}$
L.~Agostino,$^{12,}$\cite{j} V.~Arena,$^{12}$ G.~Boca,$^{12}$
G.~Bonomi,$^{12,}$\cite{k} G.~Gianini,$^{12}$ G.~Liguori,$^{12}$
M.~Merlo,$^{12}$ D.~Pantea,$^{12,}$\cite{l} S.~P.~Ratti,$^{12}$
C.~Riccardi,$^{12}$ I.~Segoni,$^{12,}$\cite{m} L.~Viola,$^{12}$
P.~Vitulo,$^{12}$ H.~Hernandez,$^{13}$ A.~M.~Lopez,$^{13}$
H.~Mendez,$^{13}$ L.~Mendez,$^{13}$ A.~Mirles,$^{13}$
E.~Montiel,$^{13}$ D.~Olaya,$^{13,}$\cite{n} A.~Paris,$^{13}$
J.~Quinones,$^{13}$ C.~Rivera,$^{13}$ W.~Xiong,$^{13}$
Y.~Zhang,$^{13,}$\cite{o} J.~R.~Wilson,$^{14}$ K.~Cho,$^{15}$
T.~Handler,$^{15}$ D.~Engh,$^{16}$ M.~Hosack,$^{16}$
W.~E.~Johns,$^{16}$ M.~Nehring,$^{16,}$\cite{p} P.~D.~Sheldon,$^{16}$
K.~Stenson,$^{16}$ M.~Webster,$^{16}$ and M.~Sheaff$^{17}$}  

\address{( The FOCUS Collaboration )\\
$^{1}$ University of California, Davis, CA 95616, USA\\
$^{2}$ Centro Brasileiro de Pesquisas F\'isicas, Rio de Janeiro, RJ, Brazil\\
$^{3}$ CINVESTAV, 07000 M\'exico City, DF, Mexico\\
$^{4}$ University of Colorado, Boulder, CO 80309\\
$^{5}$ Fermi National Accelerator Laboratory, Batavia, IL 60510\\
$^{6}$ Laboratori Nazionali di Frascati dell'INFN, Frascati, Italy, I-00044\\ 
$^{7}$ University of Illinois, Urbana-Champaign, IL 61801\\
$^{8}$ Indiana University, Bloomington, IN 47405\\
$^{9}$ Korea University, Seoul, Korea 136-701\\
$^{10}$ INFN and University of Milano, Milano, Italy\\
$^{11}$ University of North Carolina, Asheville, NC 28804\\
$^{12}$ Dipartimento di Fisica Nucleare e Teorica and INFN, Pavia, Italy\\
$^{13}$ University of Puerto Rico, Mayaguez, PR 00681\\
$^{14}$ University of South Carolina, Columbia, SC 29208\\
$^{15}$ University of Tennessee, Knoxville, TN 37996\\
$^{16}$ Vanderbilt University, Nashville, TN 37235\\
$^{17}$ University of Wisconsin, Madison, WI 53706 \\}

\date{\today}
\maketitle
\begin{abstract}
We have studied hadronic four-body decays of $\DP$ and $\DS$ mesons 
with a $\KS$ in the final state using data recorded during the
1996-1997 fixed-target run at Fermilab high energy photoproduction
experiment FOCUS. We report a new branching ratio measurement of
$\Gamma(\DP\RAW\ws)/\Gamma(\DP\RAW\ppp)=0.0768\pm0.0041\pm0.0032$. We
make the first observation of three new decay modes with
branching ratios
$\Gamma(\DP\RAW\rs)/\Gamma(\DP\RAW\ppp)=0.0562\pm0.0039\pm0.0040$,
$\Gamma(\DP\RAW\kkp)/\Gamma(\DP\RAW\ppp)=0.0077\pm0.0015\pm0.0009$, and
$\Gamma(\DS\RAW\rs)/\Gamma(\DS\RAW\ws)=0.586\pm0.052\pm0.043$, where
in each case the first error is statistical and the second error is
systematic.  
\end{abstract}
\twocolumn
}
To understand hadronic decays of a heavy quark system one needs to
address final state interactions. These become more complicated as the
number of final state hadrons increase, since the signatures of the
weak decay of the heavy quark are masked by the hadronic degrees of
freedom. Theoretical predictions are still limited to two-body decays,
which have been analyzed extensively in the theoretical
literature~\cite{REF:bigi,REF:Wise,REF:Orr}. For example, Bauer, Stech
and Wirbel~\cite{REF:Bauer} have used a factorization
approach. Bedaque, Das, and Mathur~\cite{REF:Bedaque} and Kamal, Verma
and Sinha~\cite{REF:Kamal} have used heavy quark effective
theory. Much less is known about four-body hadronic decays than about
two or three-body decays. More experimental data on higher
multiplicity modes is essential to improving our understanding of the
decay process in heavy quark systems. In this letter we report on
$\DP$ and $\DS$ branching ratios into four-body final states involving
a $\KS$. We measure the $\DP$ decay rates into $\ws$, $\rs$, and
$\kkp$ relative to $\ppp$ and the decay rate of $\DS\RAW\rs$ relative
to $\DS\RAW\ws$ (throughout this letter the charge conjugate state is
implied). Among these final states only the $\ws$ final state has been
observed previously~\cite{REF:ARGUS}. These final states have several
interesting resonant contributions which will be the subject of a
future report.       

We collected the data for this study during the 1996-1997 fixed-target
run of the photoproduction experiment, FOCUS at Fermilab. This
experiment utilized a forward multi-particle spectrometer to study
charmed particles produced by interactions of high energy photons,
$\langle E_{\gamma} \rangle \approx$ 180 GeV, with a segmented BeO
target. Charged particles were traced by silicon microstrip vertex
detectors. These detectors provide excellent separation between the
reconstructed production and decay vertices. The vertex resolution is
approximately 6 $\mu$m in the transverse direction and 300 $\mu$m in
the longitudinal direction. Three multi-cell threshold \Ckov detectors
were used to identify charged hadrons.        

We reconstructed the $\KS$ candidates using the decay
$\KS\RAW\pi^+\pi^-$ and calculated the error on the $\KS$ mass for
each candidate. We required the $\KS$ mass be within 3$\sigma$ of the
nominal $\KS$ mass which rejects nearly all of the $e^+e^-$ pair
background. To reduce the $\LMB\RAW p\pi^-$ background we applied a
\Ckov particle identification cut of 5 units on the difference in the
log likelihoods between the pion and proton hypothesis~\cite{REF:JIM}
on the $\KS$ daughter track with higher momentum.
  
We selected the final states using a candidate driven vertexing
algorithm~\cite{REF:SVERT}. A secondary vertex was formed from the
reconstructed tracks and the momentum vector of the charm candidate
was used as a {\em seed} track in finding the primary vertex in the
event. We required that the primary and secondary vertices be formed
with a confidence level greater than 1\%. The significance of
separation between the primary and secondary vertex is called
$\ell/\sigma_\ell$. We required $\ell/\sigma_\ell$ $>$ 9 for $\DP$
candidates and $\ell/\sigma_\ell$ $>$ 7 for $\DS$ candidates. We also
required the proper decay time be less than 5 times the candidate
particle's lifetime\cite{REF:PDG}. The vertexing algorithm provides
two estimators of the relative isolation of the vertices. The {\em
primary vertex isolation} (ISO1) estimator is the confidence level of
the hypothesis that a track in the secondary vertex was also in the
primary vertex; the {\em secondary vertex isolation} (ISO2) is the
confidence level that another track originates from the secondary
vertex. We required ISO1 be less than 1\% and ISO2 less than
0.1\%. For the $\DP\RAW\ws$ decay mode, this removes 80\% of the
background and retains 70\% of the signal. Further, we required that
the decay vertices be out of target material, eliminating backgrounds
from interactions which are induced by particles from the primary
interaction or from conversions of spurious photons.       

In addition to the vertexing requirements, we used \Ckov particle
identification cuts. For the charged kaon candidates we required the
kaon hypothesis be favored over the pion hypothesis by 2 units of
likelihood (for the $\DP\RAW\kkp$, we required 2 units for the faster
kaon but only 1 unit for the slower kaon).

Finally, the charm candidates were required to have a momentum greater
than 30 GeV/$c$, which removed combinatoric non-charm backgrounds. For
all decay modes we selected cuts by maximizing the figure of merit
defined as ${\mathcal N}^2_{\rm signal}/({\mathcal N}_{\rm
signal}+{\mathcal N}_{\rm background})$, as well as minimizing
reflection background.            

\begin{figure}[t]
\centerline{\psfig{figure=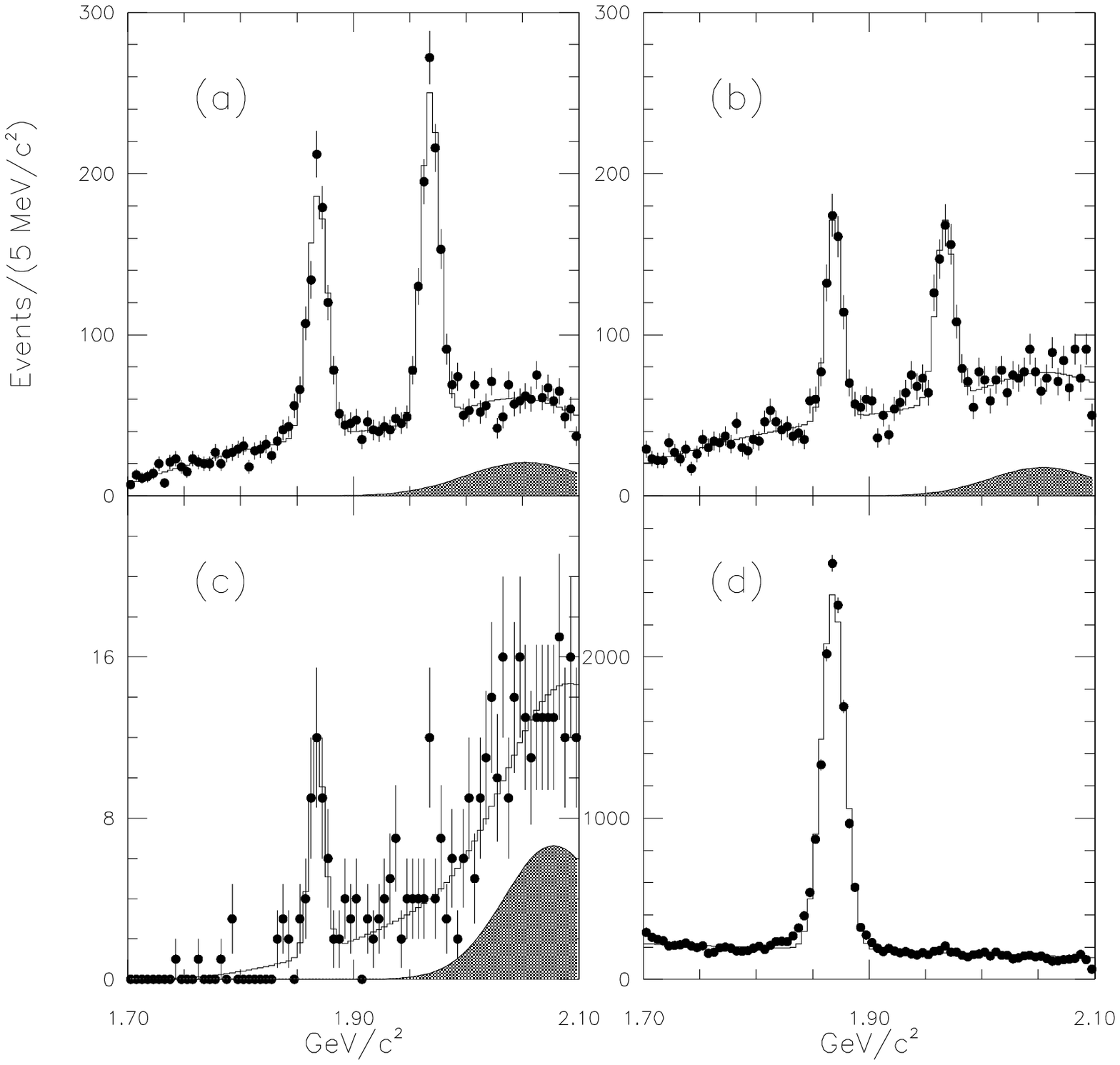,width=3.5in}}
\caption{The invariant mass distributions with $\DP$ selection
cuts. (a) shows the final state $\ws$, (b) $\rs$, (c) $\kkp$ and
(d) $\ppp$. The data are indicated by points with error bars and the
solid line is the fit. The shaded regions are charm reflection
backgrounds.}   
\label{FIG:IMDP}
\end{figure}

\begin{figure}[t]
\centerline{\psfig{figure=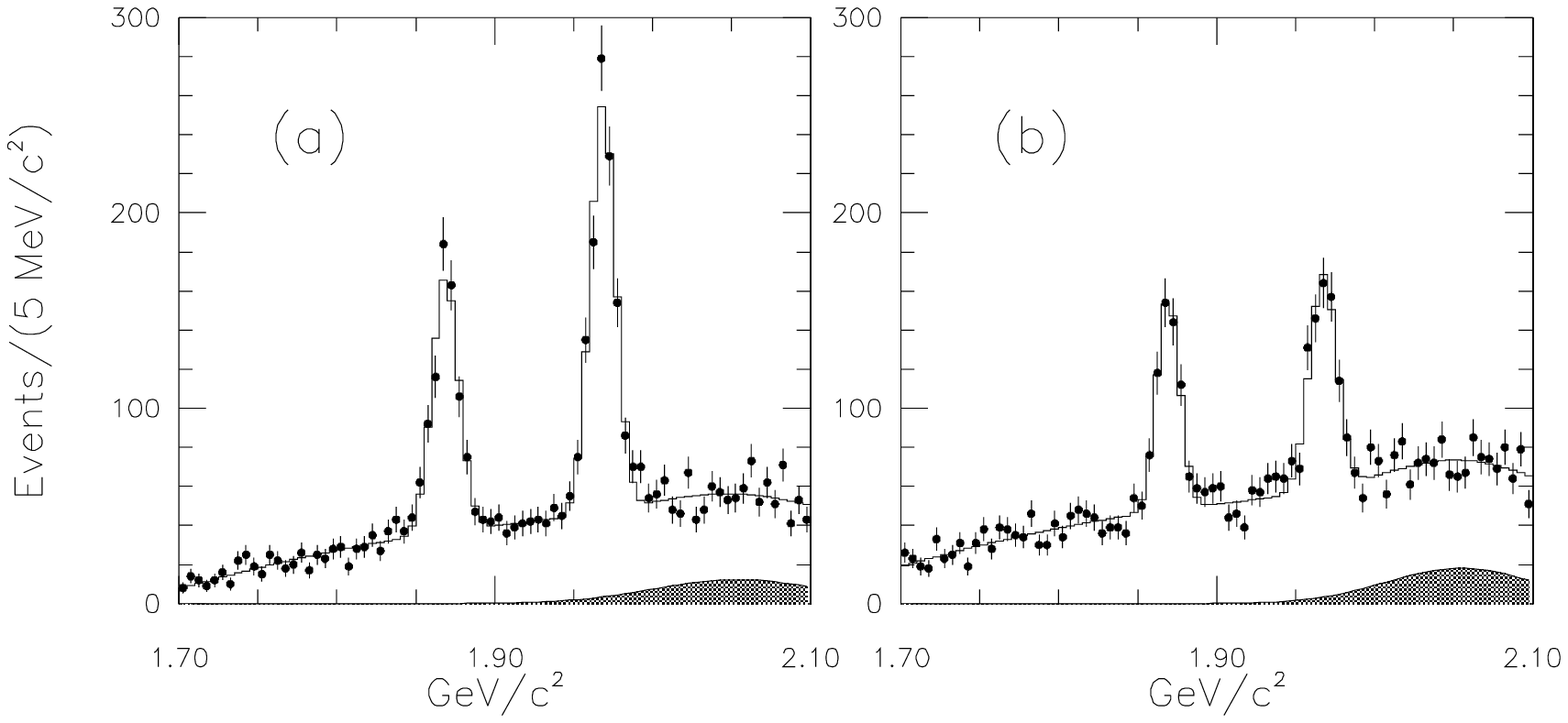,width=3.5in}}
\caption{The invariant mass distribution with $\DS$ selection
cuts. (a) shows the final state $\ws$ and (b) $\rs$. The data  are
indicated by points with error bars and the solid line is the fit. The
shaded regions are charm reflection backgrounds.}    
\label{FIG:IMDS}
\end{figure}

We investigated several possibilities for contamination of either
signal region due to reflections from partly reconstructed charmed
particles. We find no evidence in the data for an enhancement due to
$\Lambda^+_c\RAW\KS p \pi^+\pi^-$ where the $p$ is misidentified as a
$K$ or due to $\Lambda^+_c\RAW\LMB \pi^+ \pi^+ \pi^-$ where the $\LMB$
is misidentified as a $\KS$. We also investigated contamination of the
$\DS\RAW\ws$ and $\DS\RAW\rs$ signal region due to $\pi$/$K$
misidentification of the decay $\DP\RAW\ppp$. This reflection is
incorporated in the fit (see Figs.~\ref{FIG:IMDP} and
~\ref{FIG:IMDS}). Double misidentification between the $\ws$ and $\rs$
decay modes was found to be negligible.     

In order to minimize systematic effects we chose normalization channels
which have similar topologies to the signal modes, and we applied the
same vertex requirements. Fig.~\ref{FIG:IMDP} shows the invariant mass
distributions using the $\DP$ selection cuts and Fig.~\ref{FIG:IMDS}
shows the invariant mass distributions using the $\DS$ selection
cuts. We parameterized the signal with a Gaussian and the background
with a quadratic polynomial plus a reflection shape. For the
$\DP\RAW\ppp$ normalization mode we used a linear background. The
reflections in Figs.~\ref{FIG:IMDP}(a), \ref{FIG:IMDP}(b) and
Figs.~\ref{FIG:IMDS}(a), \ref{FIG:IMDS}(b) from $\DP\RAW\ppp$ and
Fig.~\ref{FIG:IMDP}(c) from $\DP\RAW\ws(\rs)$ occur due to the
misidentification of a $\pi$ as a $K$. The shapes were determined by
Monte Carlo simulations and the reflection amplitudes are free
parameters in the fit.           


The final states in this study have resonant substructure in their
decay modes which may affect the reconstruction efficiency. We
generated a Monte Carlo for each final state using an incoherent mix
of various sub-resonant decay modes and a non-resonant
contribution. For the decay $\DP\RAW\ppp$ we used its known sub-decay
modes to determine the efficiency. Mini-Monte Carlo studies showed
this efficiency is consistent with that of allowing interferences
between the states. For the other decay modes, where little is known
about the resonant substructure, we determined the efficiencies based
on which sub-decay we found to be dominant. The decay mode used for
the $\ws$ ($\kkp$) final state was $\KSTB K^{*+}$
($\KS\phi\pi^+$). The $\rs$ final state has contributions from many
sub-decay modes; therefore we assumed a non-resonant substructure to
obtain the efficiency for this final state. With these Monte Carlo
efficiencies, we measured the relative branching ratios which are
summarized in Table \ref{TABLE1}. For each of the signal modes we also
obtained efficiencies with different resonant and non-resonant
substructure combinations. We used the spread of the various resonant
and non-resonant efficiencies  to determine the systematic error on
the efficiency calculation for each state. We estimated the relative
uncertainties in efficiencies due to uncertainties in the branching
fractions into resonant sub-states to be 3.9\% for $\DP\RAW\ws$, 6.7\%
for $\DP\RAW\rs$, 9.8\% for $\DP\RAW\kkp$, 4.6\% for $\DS\RAW\ws$, and
4.6\% for $\DS\RAW\rs$.        

A study of the stability and behavior for each branching ratio was
performed using variations of our analysis cuts; we found no bias from
the choice of analysis cuts. Further, we split our data into independent
subsamples based on $D$ momentum and the different run periods in
which the data were accumulated. This technique is described in detail
in reference~\cite{REF:DAN}. We found no systematic uncertainties from
splitting our data.     

The systematic uncertainty for each branching ratio includes
efficiency dependencies from sub-resonant states and from variation of
the fit parameters (mostly due to background parameterization
variation).

\begin{table}
\newcolumntype{Y}{>{\centering\arraybackslash}X}%
\begin{tabularx}{\linewidth}%
		{>{\setlength{\hsize}{1.15\hsize}}Y%
		 >{\setlength{\hsize}{0.55\hsize}}Y%
		 >{\setlength{\hsize}{1.15\hsize}}Y} \hline \hline
Decay Mode   &${\mathcal N}_{\rm signal}$&$\Gamma_{\rm rel}$ \\ \hline \hline
$\DP\RAW\ws$ &670  $\pm$35   &0.0768$\pm$0.0041$\pm$0.0032 \\   
$\DP\RAW\rs$ &469  $\pm$32   &0.0562$\pm$0.0039$\pm$0.0040 \\   
$\DP\RAW\kkp$&35   $\pm$7    &0.0077$\pm$0.0015$\pm$0.0009 \\ 
$\DP\RAW\ppp$&11590$\pm$121  &1                            \\ \hline   
$\DS\RAW\rs$ &476  $\pm$36   &0.586$\pm$0.052$\pm$0.043 \\ 
$\DS\RAW\ws$ &837  $\pm$38   &1                          \\ \hline \hline  
\end{tabularx} 
\vspace{0.3cm}
\caption{Measured branching ratios.  $\Gamma_{\rm rel}$ is the branching 
ratio relative to $\DP\RAW\ppp$
for the $\DP$ modes and $\DS\RAW\ws$ for the $\DS$ modes. The errors on 
the branching ratios are statistical and systematic, respectively.}    
\label{TABLE1}
\end{table}

To conclude, we have improved the previous measurements of the
$\DP\RAW\ws$ branching ratio and have made the first observation of
the $\DP(\DS)\RAW\rs$ and $\DP\RAW\kkp$ decay processes using a high
statistics sample of photoproduced charmed particles from the FOCUS
experiment at Fermilab.           
  
We wish to acknowledge the assistance of the staffs of Fermilab and
the INFN of Italy, and the physics departments of the collaborating
institutions. This research was supported in part by the
U. S. National Science Foundation, the U. S. Department of Energy, the 
Italian Istituto Nazionale di Fisica Nucleare and Ministero
dell'Universit\`a e della Ricerca Scientifica e Tecnologica, the
Brazilian Conselho Nacional de Desenvolvimento Cient\'ifico e
Tecnol\'ogico, CONACyT-M\'exico, the Korean Ministry of Education, and
the Korean Science and Engineering Foundation.

\end{document}